\documentclass[final, 5p, times, twocolumn]{elsarticle}
\usepackage{lineno}
\usepackage{times}
\usepackage{graphicx,subfigure,epsfig,overpic,color}
\usepackage{url,amsmath,amssymb,amsthm,bm}
\usepackage{multirow,proof}
\usepackage{threeparttable,array,tabularx}

\usepackage[ruled,boxed]{algorithm2e}

\journal{X}

\begin{document}

\begin{frontmatter}

\title{Towards Scalable Topic Detection on Web via Simulating L\'{e}vy Walks Nature of Topics in Similarity Space}

\author[a]{Junbiao Pang}
\ead{junbiao\_pang@bjut.edu.cn}

\author[a,b]{Qingming~Huang}
\ead{qmhuang@ucas.ac.cn}

\address[a]{Faculty of Information Technology, Beijing University of Technology, No.100 Pingleyuan Road, Chaoyang District, Beijing 100124, China}

\address[b]{School of Computer and Control Engineering, University of Chinese Academy of Sciences,\\ No.19 Yuquan Road, Shijingshan District, Beijing 100049, China}

\begin{abstract}

Organizing a few webpages from social media websites into popular topics is one of the key steps to understand trends on web. Discovering popular topics from web faces a sea of noise webpages which never evolve into popular topics. In this paper, we discover that the similarity values between webpages in a popular topic contain the statistically similar features observed in L\'{e}vy walks. Consequently, we present a simple, novel, yet very powerful Explore-Exploit (EE) approach to group topics by simulating L\'{e}vy walks nature in the similarity space. The proposed EE-based topic clustering is an effective and efficient method which is a solid move towards handling a sea of noise webpages. Experiments on two public data sets demonstrate that our approach is not only comparable to the state-of-the-art methods in terms of effectiveness but also significantly outperforms the state-of-the-art methods in terms of efficiency.

\end{abstract}

\begin{keyword}

User-Generated Content, Web Topic Detection, L\'{e}vy Walks, Explore-Exploit, Noise Robust Clustering

\end{keyword}

\end{frontmatter}


\section{INTRODUCTION}\label{sec:intro}

With the rapid development of social media, User-Generated Content (UGC)~\cite{pang2013unsupervised} is quite pervasive for people to share their options and experiences on websites. As a result, the unprecedented explosion in the volume of UGC data has made it difficult for web users to quickly access popular contents~\cite{shahaf2010connecting}. Topic detection on web~\cite{pang2013unsupervised} is such an effort to organize web data into more meaningful popular topics automatically. Essentially, topic detection on web is like looking for a needle in the haystack, \emph{i.e.}, a tiny fraction of interesting webpages from a large amount of UGC data is organized into a seminal event~\cite{pang2013unsupervised}.

The task of topic detection on web is totally different from Topic Detection and Tracking (TDT)~\cite{Allan-Carbonell-Doddington-Yamron-Yang-1998} which aims to assign each news article into at least one topic. Consequently, TDT organizes topics for professionally edited articles~\cite{Allan-kuwer-00} which are totally different from User-Generated Content (UGC) on social media as follows: 1) the textual and the visual information from social media tends to be short, sparse and noisy~\cite{zhao-ecir-2011} due to the less constraint on social media; and 2) only a small fraction of webpages belongs to popular topics, for example, about 1\% webpages are organized into popular topics in Youtube~\cite{pang2013unsupervised}. Therefore, topic detection on web not only faces the inefficient features but also has to handle a large number of noise webpages.

The key problem of topic detection on web is how to organize ``hot'' topics confronting with a sea of noise webpages, when only inefficient features are extracted from the noisy and sparse multi-media data. A naive approach is to cluster web topics in the presence of noise webpages. However, a sea of noises in UGC data would overwhelm the conventional approaches~\cite{bojchevsik-robust-spectral-kdd17}~\cite{maurus-skinny-dip-kdd16} which are proposed to handle the presence of the a small number of noises.

In order to remove the adverse effects from a sea of noise webpages, the conventional methods~\cite{Zhang-Li-Chu-Wang-Zhang-Huang-2013}~\cite{pang2013unsupervised} adopt a seemingly reasonable assumption that a hot topic should be a dense clustering. That is, the intra-similarities between two webpages from the same topic should be larger than the inter-similarities. However, this assumption barely holds for hot topics on web due to the following two problems:
\begin{itemize}
\item []\textbf{1) Sparse and Noisy UGC Data Result in Inefficient Features.} UGC data are posted with few constraints on social media websites. Traditional features are computationally simple but not sufficiently efficient for webpages on social media, such as, Term Frequency-Inverse Document Frequency (TF-IDF), word embedding~\cite{penn-14-emNLP-glove} for long texts. Although deep word embedding based on Transformer~\cite{ashish-nips-2017} (\emph{i.e.}, BERT,~\cite{devlin-etal-2019-bert}) are efficient but are computationally intensive for web topic detection; 
\item []\textbf{2) Semantic Gap between Low-level Features and High-level Semantics.} Low-level features are difficult to be correlated with the high-level semantics. Therefore, a larger similarity value between two webpages with a certain chance cannot indicate that these two webpages are more semantically similar.
\end{itemize}

In this paper, we seek a \emph{lightweight}, and \emph{optimization-free} approach to generate web topics in the presence of both~\emph{inefficient features} and~\emph{a sea of noises}, based on four motivations. Firstly, the structures and contents of web topics are significantly diverse from each other; therefore, a lightweight feature extraction is expected to avoid the time-consuming optimization. Secondly, the optimization-free method is desired to handle a large-scale dataset~\cite{pang-tao-lpd-icme-2016} and achieve a good generalization ability. Thirdly, we want to avoid the open problem about encoding efficient features for short texts~\cite{liu-icml-2009} with complex encoding method~\cite{devlin-etal-2019-bert}. Fourthly, detecting ``hot'' topics on web has been confronted with a sea of noises. In summary, we desire a computationally simple yet efficient method to leverage the nature of web topics.

This paper studies the statistical pattern of web topics in the similarity space, finding that a web topic has a similar feature which is the hallmark of L\'{e}vy walks~\cite{perkins-glass-edwards-nature-comunication-2014}. Concretely, if we draw an analogy between the similarities between webpages and the flights in L\'{e}vy walks, the distribution of the similarities in a hot topic closely matches the heavy-tailed distribution which are typically used to characterize the flights in L\'{e}vy walks.

L\'{e}vy walks~\cite{viswanathan-levy-walk-physica-a-2002}~\cite{rhee-shin-levy-walk-TN-2011} are random walk models, where the step-lengths have a heavy-tailed probability distribution. A~\emph{flight} in L\'{e}vy walks is defined as a step-length of a particle from one location to another without a directional change. Intuitively, L\'{e}vy walks consist of many short flights and a few exceptionally long flights that eliminate the effect of such short flights. Therefore, L\'{e}vy walks have been found to describe the mobility patterns of foraging animals such spider monkey~\cite{viswanathan-levy-walk-physica-a-2002}.

When L\'{e}vy walks are used to organize web topics, the key questions are therefore: 1) How to simulate the exceptional flights in L\'{e}vy walks with unknown parameters, 2) how to determine the number of requisite flights to organize topics, and 3) whether such a walk leads to any advantages in organizing topics. Motivated by the above three questions,
\begin{itemize}
\item []1. This paper proposes an Explore-Exploit (EE) approach to simulate L\'{e}vy walks in the similarity space, elegantly avoiding unknown parameters for different L\'{e}vy walks.
\item []2. A multi-thresholds approach is proposed to terminate the growth of web topics, producing a set of overlapped topics. Therefore, the recall of a topic detection system is naturally guaranteed.
\item []3. One important technique component of this paper is Poisson Deconvolution (PD) which is a state-of-the-art approach to rank the interestingness of topics~\cite{pang2013unsupervised}~\cite{pang-tao-lpd-icme-2016}~\cite{pang-tao-neurocomputing-2018} for topic detection on web. PD empirically shows that the precise choice of the number of topics is not critical: a small number of hot topics tend to be correctly ranked ahead. Therefore, the precision of a detection system is expected to be guaranteed.
\end{itemize}

To the best of our knowledge, this paper is the first to discover the similar features between L\'{e}vy walks and web topics in the similarity space, presenting a comprehensive series of experiments to illustrate the benefits of this novel observation for topic detection on web. The proposed method is computationally simple, yet exceptionally powerful. Simply by selecting a set of webpages by the proposed EE approach, with no further parameter tuning, we find a new way to organize web topics that is comparable to the current state-of-the-art in terms of effectiveness, but significantly surpasses the state-of-the-art in terms of efficiency.

The rest of this paper is organized as follows: Section~\ref{sec:relatedwork} reviews the related work. We describe the details of our approach in Section~\ref{sec:method}. Experimental results are presented in Section~\ref{sec:experiments} and the paper is concluded in Section~\ref{sec:conclusion}.

\section{Related Work}\label{sec:relatedwork}

It is seemingly intuitive to assume that any two webpages in a topic should have higher similarities than the other two ones, although this assumption is empirically verified to be wrong in Section~\ref{sec:sub:nature-of-topic}. Based on the assumption above, existing approaches often organize webpages into topics by clustering.

\subsubsection{Clustering in Sea of Noises}

The most popular way to define a cluster is to calculate average intra-similarity within a topic. However, clustering in sea of noises challenges these traditional methods~\cite{TSMCS-DBSCAN-21}~\cite{ACM-data-density-23}. Therefore, Rather than the partition-based method, clustering from a seed point largely avoids to introduce noises into clustering. For example, Pang et al.~\cite{pang2013unsupervised} used maximal cliques on a graph as topics. Cao et al.~\cite{Cao-Ngo-Zhang-Li-2011} generated events on video tags by $k$-means based on textual-visual similarity. Zhang et al.~\cite{Zhang-Li-Chu-Wang-Zhang-Huang-2013} proposed to use Graph Shift (GS)~\cite{Liu-Yan-2010} to find dense subgraphs as topics. The clustering from a seed point usually discover a small number of hot topics, \emph{i.e.}, recall would be relatively low. Because finding the promising seeds of clusters is difficult in present of sparse and noisy data which widely occur in social media.

Instead of calculating the intra-similarity~\cite{Cao-Ngo-Zhang-Li-2011} \cite{he-pami-10}, some approaches adopt advanced clustering algorithms. In~\cite{xu-liu-gong-sigir-03}, Nonnegative Matrix Factorization (NMF) showed more accurate performance than that of the spectral methods in document clustering. Recently, topic models have been proposed to infer hidden themes for document analysis, including Hierarchical Dirichlet Processes (HDP)~\cite{teh-jmsa-06}, and various variations. These topic models generally work well on long documents~\cite{baldwin-CoNLL-2009}~\cite{wang-etal-2023-text2topic}. However, these approaches are not directly extensible for clustering topic in a sea of noises.

\subsubsection{Hot Topic Detection on Web}

In fact, web topics are not equivalent to clusters since the proportion of the noisy and sparse data is larger than that of other scenarios, \emph{e.g.}, high-dimension data~\cite{pandove-goel-rani-review-high-dimension-18}, budgeted optimization~\cite{byrka-pensyl-rybicki-srinivasan-trinh-budgeted-17}. Therefore, detection-as-ranking approach is proposed by PD~\cite{pang2013unsupervised}. PD generates over-complete topics by off-the-shelf clustering methods~\cite{yang2012clustering}, and then detects ``hot'' topics by ranking the interestingness of topics. However, generating over-complete topics for PD is a time-consuming process~\cite{pang2013unsupervised}~\cite{pang-tao-lpd-icme-2016}. The proposed EE-based method is a solid move forward making PD scale up to large-scale data sets.



L\'{e}vy walks has been found to describe the foraging behaviors of animals~\cite{viswanathan-levy-walk-physica-a-2002}. Because L\'{e}vy walks are prominent random strategy where many short flights are occasionally alternated with a long one. Therefore, L\'{e}vy walks provide a more efficient method to visit sites~\cite{li-reis-levy-walk-prl-2010}. The other important application of L\'{e}vy walks is the routing in a network~\cite{perkins-glass-edwards-nature-comunication-2014}. To the best of our knowledge, this paper is the first to bridge the gap between L\'{e}vy walks and web topics.

\section{Simulating L\'{e}vy-Walk Nature of Web Topics in Similarity Space}\label{sec:method}

\begin{figure*}[t!]
  \centering
  \subfigure[Exponentiated Weibull]{
    \label{fig:subfig:topic1} 
    \includegraphics[width=.236\textwidth]{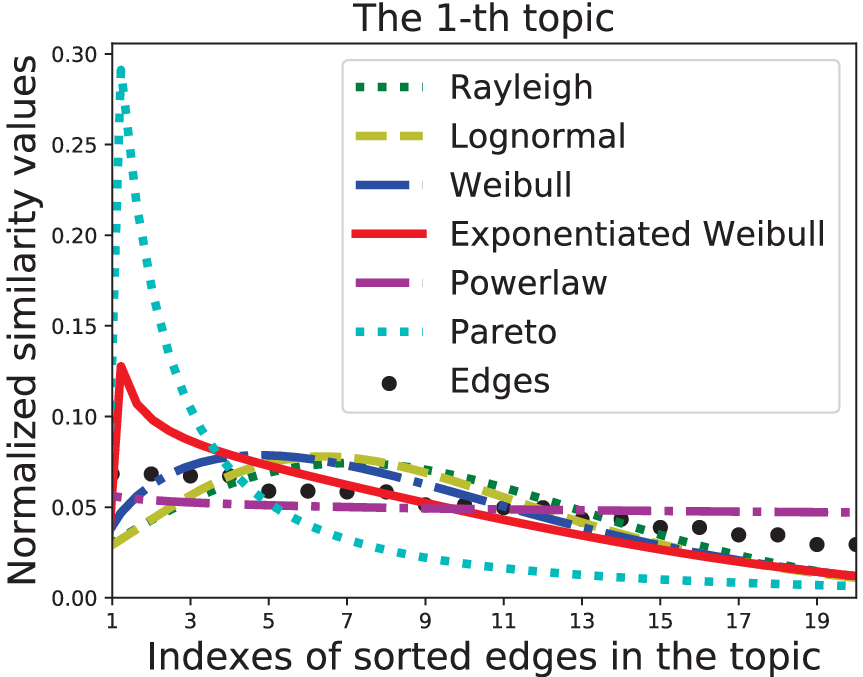}}
  \subfigure[Power law]{
    \label{fig:subfig:topic2} 
    \includegraphics[width=.236\textwidth]{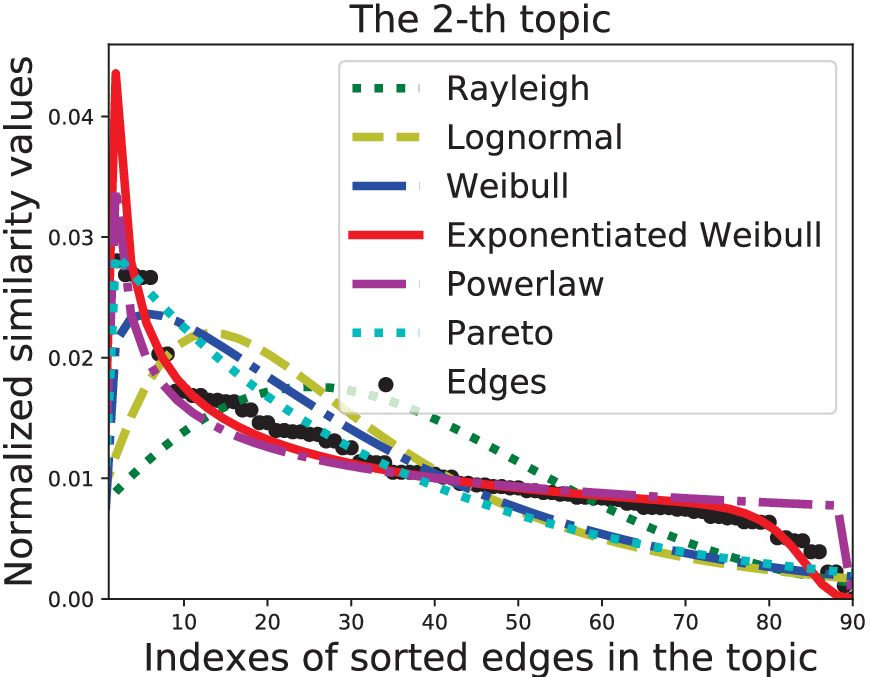}}
     \subfigure[Exponentiated Weibull]{
    \label{fig:subfig:topic3} 
    \includegraphics[width=.236\textwidth]{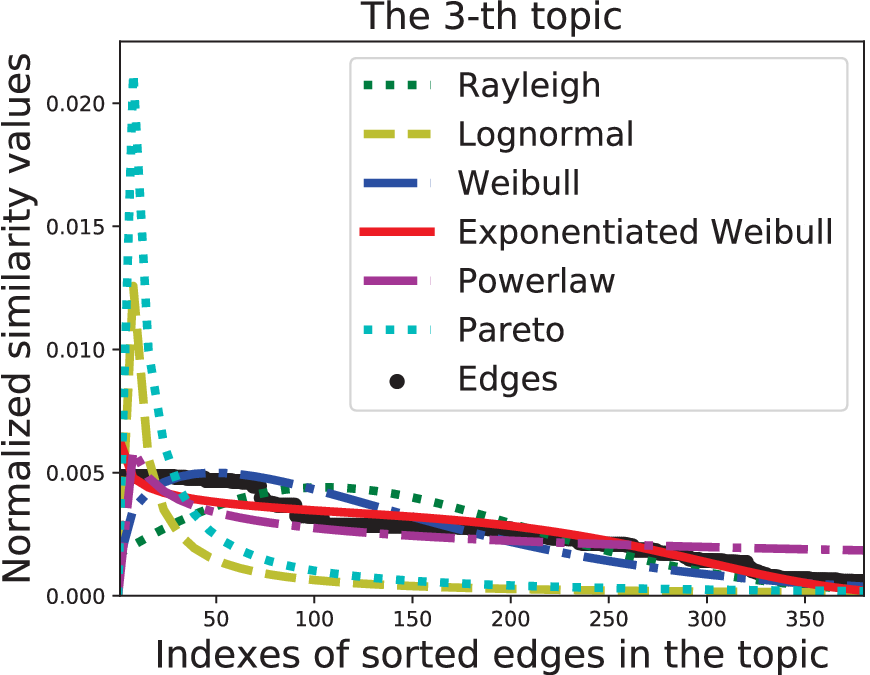}}
     \subfigure[Power law]{
    \label{fig:subfig:topic4} 
    \includegraphics[width=.236\textwidth]{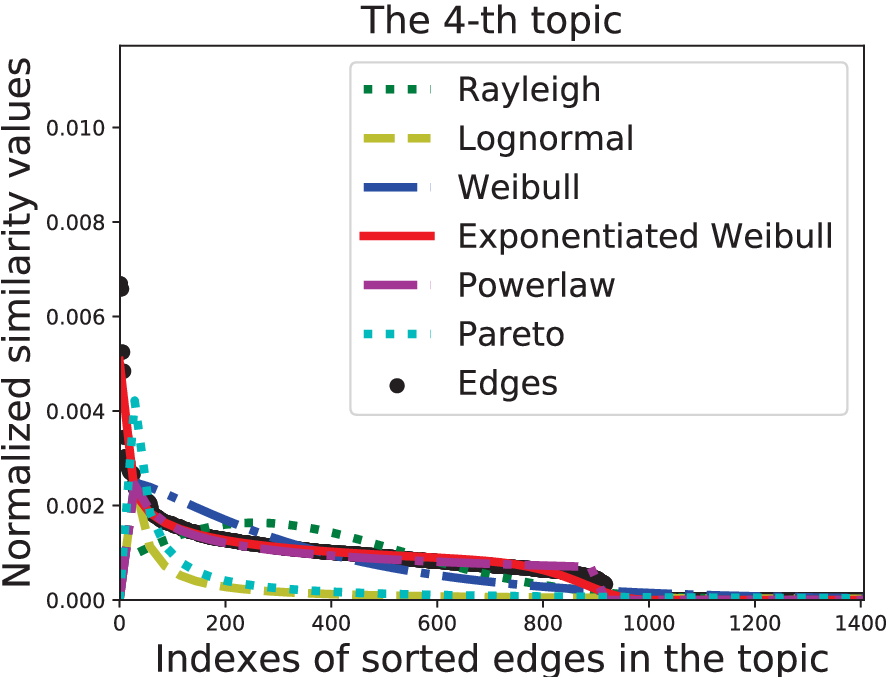}}
  \caption{The 1-th, 2-th, 3-th and 4-th topics are best fitted by exponentiated Weibull, power law, exponentiated Weibull, power law, exponentiated Weibull, respectively.}
  \label{fig:heavy-tailed-topics} 
\end{figure*}

\begin{figure*}[t!]
  \centering
  \subfigure[]{
    \label{fig:subfig:topic1-values} 
    \includegraphics[width=.236\textwidth]{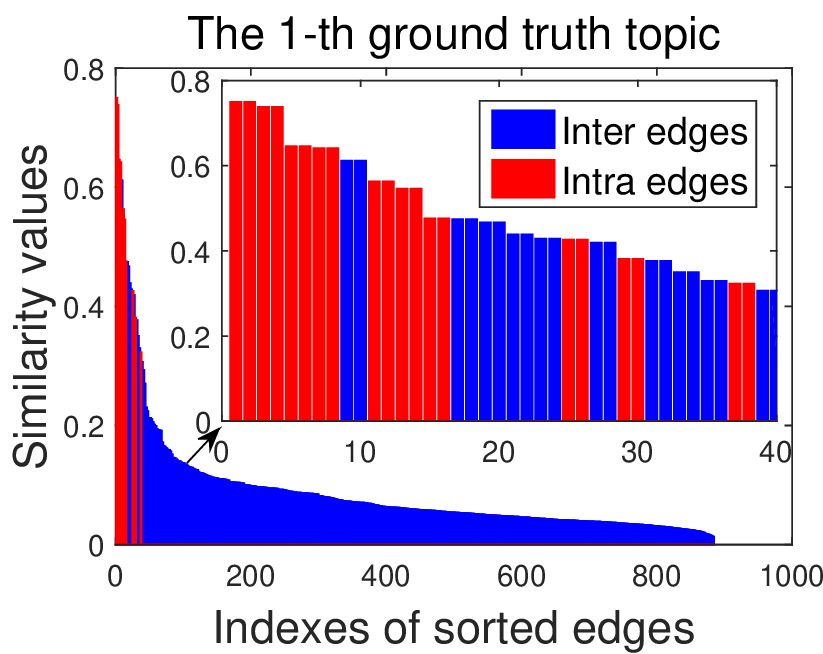}}
     \subfigure[]{
    \label{fig:subfig:topic2-value} 
    \includegraphics[width=.236\textwidth]{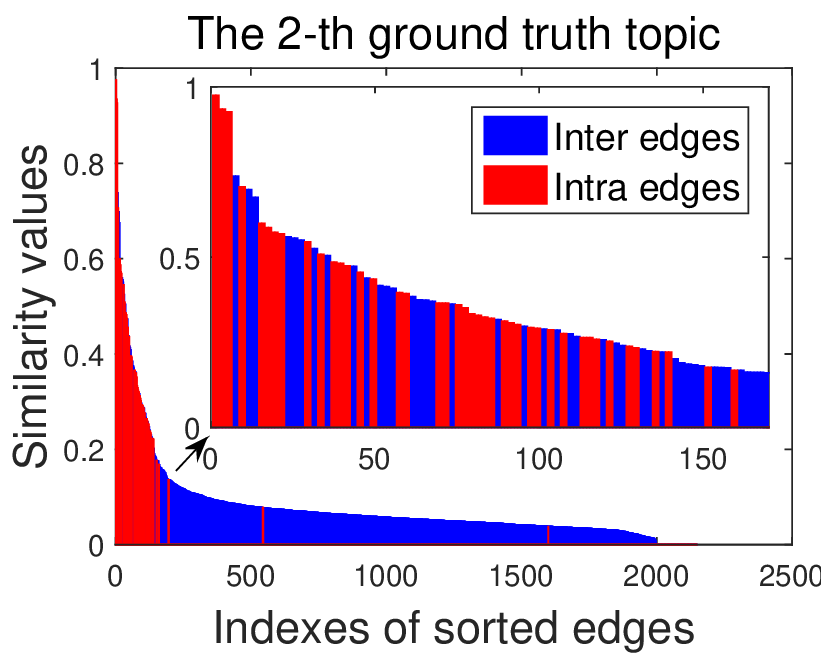}}
     \subfigure[]{
    \label{fig:subfig:topic3-value} 
    \includegraphics[width=.236\textwidth]{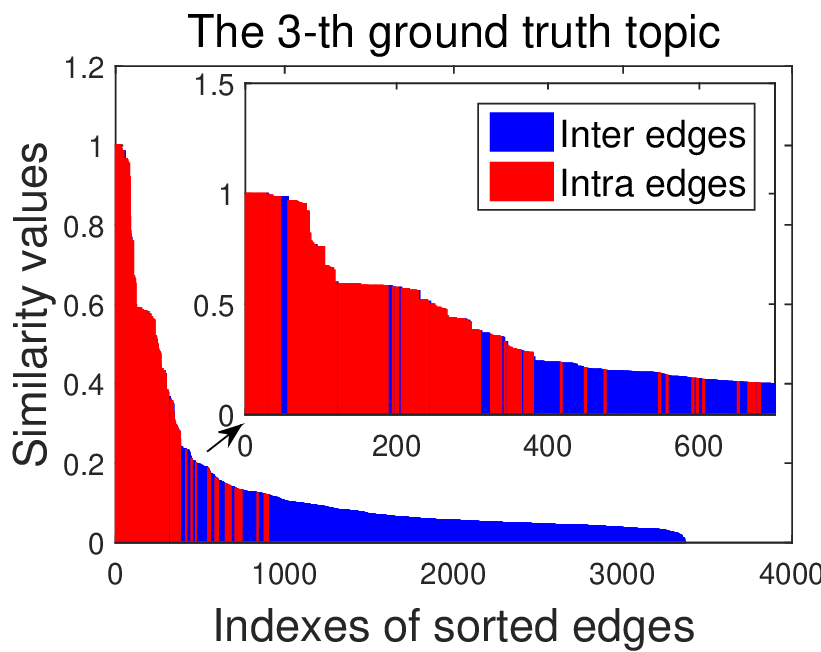}}
     \subfigure[]{
    \label{fig:subfig:topic4-value} 
    \includegraphics[width=.236\textwidth]{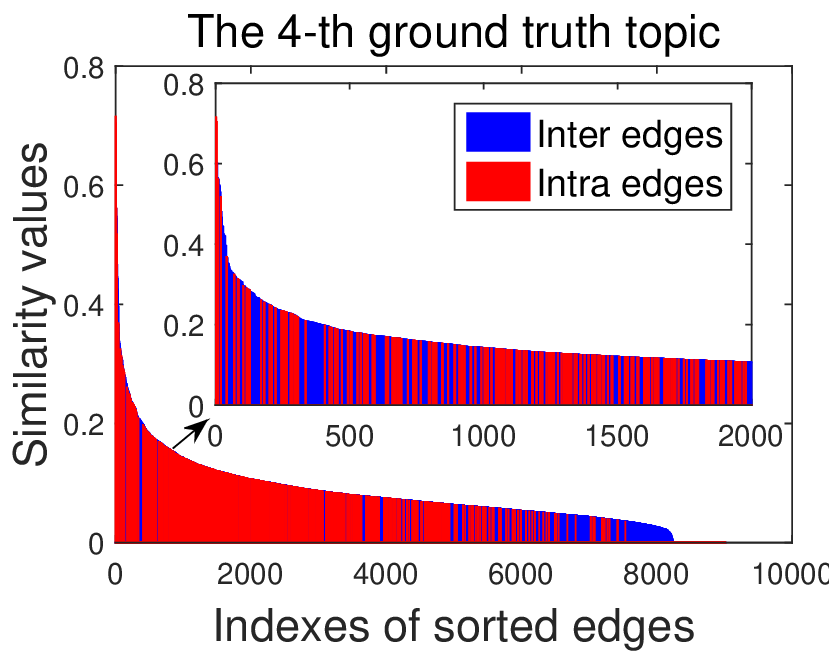}}
  \caption{The distributions of the sorted edges which include both the intra-edges and the inter-edges.}
  \label{fig:sorted-edges-topic} 
\end{figure*}

\subsection{Projecting Webpages into the Similarity Space}

Given a set of webpages $\mathcal{X}=\{\mathbf{x}_1,\ldots,\mathbf{x}_N\}$, we convert these webpages into a $k$-Nearest Neighbor ($k$-N$^2$) graph \mbox{$G=(V,E,A)$} where the $k$ nearest neighbors of each webpage are preserved. Vertex $V$ corresponds to webpages $\mathcal{X}$, the affinity matrix $A$ ($a_{ij}\in A$) corresponds to the truncated similarities between any two webpages $\mathcal{X}$, and the edge set $E$ ($e_{ij}\in E$) corresponds to the non-zero edges between any two webpages $\mathcal{X}$~\cite{pang-tao-lpd-icme-2016}. $k$ in $k$-N$^2$ graph removes some noises in the similarity space. The more noises a dataset has, the lower the parameter $k$ is. More details about how to build a $k$-N$^2$ graph can be found in our previous work~\cite{pang-tao-lpd-icme-2016}~\cite{pang2013unsupervised}.

In this paper, a webpage is simply represented by the off-the-shelf feature, \emph{i.e.}, TF-IDF, due to the following two reasons: 1) even if more advanced methods (\emph{e.g.},~\cite{liu-icml-2009}~\cite{le-mikolov-word2vec-imcl-14}) are used to represent short texts in webpages, the semantics gap is still inevitable; and 2) simpler feature means a faster processing speed.

\newtheorem{myobr}{Observation}
\newtheorem{mydef}{Definition}
\newtheorem{mythe}{Theorem}
\newtheorem{mypro}{Proposition}

\begin{mydef}[Similarity Space of a Topic on Web]
Given a $k$-N$^2$ graph $G=(V,E,A)$ and a topic $C$, the similarity space of the topic $C$ consists of the affinity values $a_{ij}$ from the edge $e_{ij}$ where at least one webpage $\mathbf{x}_i$ belongs to the topic $C$.
\end{mydef}

\begin{figure}[h!]
  \centering
  \includegraphics[width=.3\textwidth]{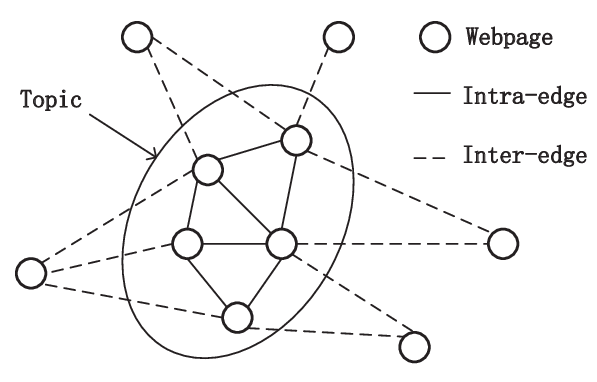}
  \caption{The similarity space of a topic consists of the affinity values from both the intra-edges and inter-edges.}
  \label{fig:similarityspace}
\end{figure}

Consequently, the edges in the similarity space could be classified into two categories (see Fig.~\ref{fig:similarityspace}): 1) \emph{Intra-edges:} the edges in the same topic; and 2) \emph{Inter-edges:} the edges across a topic. The inter-edges could be considered as the boundary of a hot topic.

\begin{table}[t!]
\centering
\small{
\caption{Well-known Heavy-tailed Distributions.}~\label{tbl:similarity-of-topic}
\begin{tabular}{|c|c|}\hline
Distribution & Probability density function (pdf) \\\hline\hline
Exponentiated Weibull$^1$ & $\left(1-\exp(-(\frac{x}{\lambda}))^k\right)^\alpha$\\\hline
Rayleigh  & $ \frac{x}{\sigma^2}\exp(\frac{-x^2}{2\sigma^2}) $\\\hline
Weibull$^2$  &   $\frac{k}{\lambda}(\frac{x}{\lambda})^{k-1}\exp\left(-(\frac{x}{\lambda})^k\right)$ \\\hline
Lognormal & $\frac{1}{x\sigma\sqrt{2\pi}}\exp\left(-\frac{\ln(x)-\mu}{2\sigma^2}\right)$ \\\hline
Power law & $a(cx)^{-k}$\\\hline
Patreto$^3$ &  $\frac{\alpha a^\alpha}{x^{\alpha+1}}$ \\\hline
\end{tabular}
\\
$x$ represents the length of flights or similarity values in this paper.
\begin{tablenotes}
\item $^1$ For $\alpha\geq1$. $^2$ When $k<1$. $^3$ For $0<a\leq x$.
\end{tablenotes}
}
\end{table}

\subsection{L\'{e}vy Walks Nature of Web Topics in the Similarity Space}\label{sec:sub:nature-of-topic}

Four ground truth hot topics from MCG-WEBV~\cite{Cao-Ngo-Zhang-Li-2011} (which will be detailed in Section~\ref{sec:experiments}) have been randomly chosen to understand web topics in the similarity space.
Concretely, the similarities of the intra-edges from the same topic are sorted in a descent order.
Fig.~\ref{fig:sorted-edges-topic} illustrates that the sorted similarities follow the heavy-tailed distributions. Two questions are naturally inspired by Fig.~\ref{fig:sorted-edges-topic} as follows:
\begin{itemize}
\item [1.] \emph{Whether all topics follow the same heavy-tailed distribution?}
\item [2.] \emph{Even when some topics follow the same distribution, whether they have the same parameters?}
\end{itemize}

To handle the questions above, we use Maximum Likelihood Estimation (MLE) to fit the sorted similarities to the well-known distributions, \emph{i.e.}, exponentiated Weibull, Rayleigh, Weibull, lognormal, Power law, and Pareto distributions. Moreover, to quantitatively find the best fitting distribution, Akaike weight $w_i$ is useful as the evidence~\cite{burnham-anderson-AIC-BIC-2004} of best fitting. Because Akaike's Information Criterion (AIC) values are greatly affected by the sample size~\cite{rhee-shin-AIC-BIC-2011}.
\begin{equation}\label{eqt:weight-AIC}
w_i=\frac{\exp(-\Delta_i/2)}{\sum_{i}^{|C|}\exp(-\Delta_i/2)},
\end{equation}
where $\Delta$ is defined as a transformation of AIC, $\Delta_i=\text{AIC}_i-\text{AIC}_{min}$,
in which $\text{AIC}_{min}$ is the minimum of different AIC values.


From Figs.~\ref{fig:heavy-tailed-topics} to~\ref{fig:sorted-edges-topic}, we have the following observations:
\begin{itemize}
\item \emph{The sorted similarities in a topic have statistically similar features as L\'{e}vy walks have.} All Figs.~\ref{fig:subfig:topic1},~\ref{fig:subfig:topic2},~\ref{fig:subfig:topic3}, and~\ref{fig:subfig:topic4} show that the similarities in a topic follow the heavy-tailed distributions.
\item \emph{Different topics follow diverse heavy-tailed distributions.} For instance, according to the Akaike weight in~\eqref{eqt:weight-AIC}, the $1$-th topic in Fig.~\ref{fig:subfig:topic1} follows the exponential Weibull distribution; while, the $2$-th topic in Fig.~\ref{fig:subfig:topic2} follows the power law distribution.
\item \emph{Some exceptional edges are contained in a hot topic.} It is seeming that all intra-edges should ranked ahead of the inter-edges. However, as illustrated in Fig.~\ref{fig:sorted-edges-topic}, some webpages are ranked ahead of the correct ones in a hot topic.
\end{itemize}

In summary, if the similarities are compared to the ``flights'' between webpages, the similarity space of a topic has two statically similar features as L\'{e}vy walks have: 1) L\'{e}vy walks and a web topic all have heavy-tailed distributions; and 2) L\'{e}vy walks  and a web topic consist of many short flights and a few exceptionally long flights. Hereafter, the two similar features above are referred to as the L\'{e}vy walks nature of a topic on web.


\subsection{Generating Topics by Simulating L\'{e}vy Walks}

We are interested in leveraging the L\'{e}vy walks nature to efficiently generate web topics. One intuitive solution is to organize some webpages into a topic by stimulating exceptionally long flights. Therefore, we should handle the following problems: 1) select seed webpages to ``grow'' a topic; and 2) inject unexceptional webpages into a topic.

\subsubsection{Selecting Important Webpages as Seeds}

We simulate the information flow from one webpage to another as a random walk process on the $k$-N$^2$ graph. Intuitively, the total information sent from one webpage to another is decided by two aspects: the visit frequency and the amount of information at each visit.

\begin{figure}[t!]
  \centering
  \includegraphics[width=.40\textwidth]{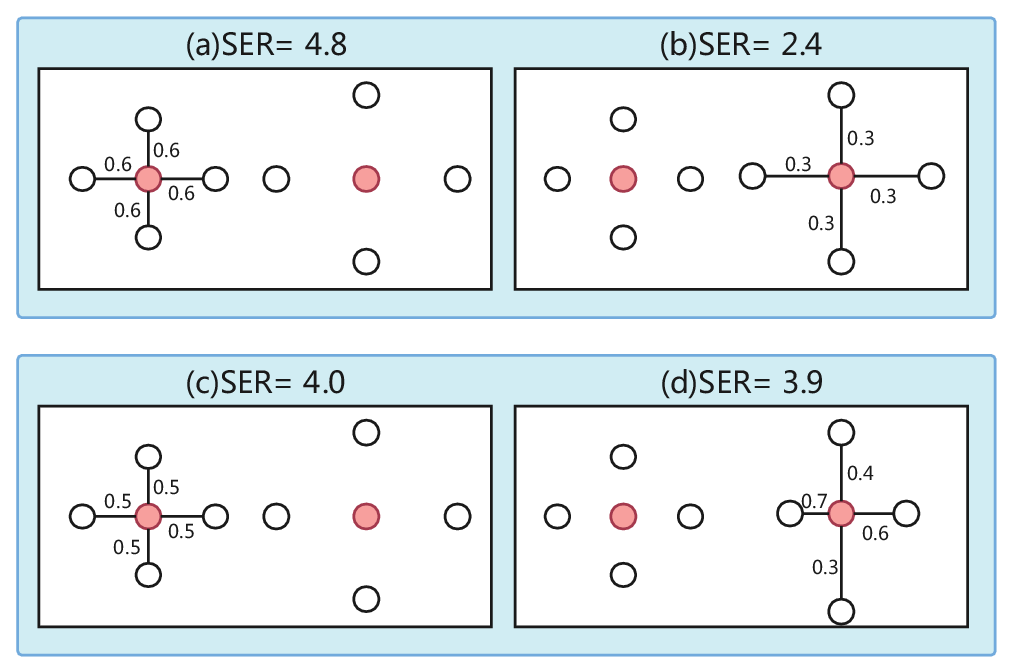}
  \caption{The role of SER of indicating compactness and homogeneousness seeds.}
  \label{fig:SER}
\end{figure}

If the finite-state Markov chain is irreducible and aperiodic, the stationary distribution $\pi$ is considered as the visit frequency of a node at the equilibrium of the random walk. Concretely, for a random walk process, the element of $\pi_i$ at webpage $i$ can be simply computed as
\begin{equation}\label{eqt:stationarydistribution}
\pi_j=\alpha\sum_{i\in V}\pi_i P_{ij} + (1-\alpha)\frac{1}{N},
\end{equation}
where $P_{ij}=A_{ij}/d_i$, $d_i=\sum_jA_{ij}$ is the out-degree of the node $i$, and $\alpha\in (0,1)$ denotes the decay factor that respects the probability with which a walker follows the connected edges.~\eqref{eqt:stationarydistribution} exactly is PageRank~\cite{page-brin-motwani-pagerank-99} algorithm.

The entropy rate qualities the uncertainty of a stochastic process $\mathcal{X}=\{X_t|t\in T\}$ where $T$ is some index set. For a discrete random process, the entropy rate is defined as an asymptotic measure $\mathcal{H}(\mathcal{X})=\lim_{t\rightarrow \infty}H(X_t|X_{t-1},X_{t-2},\ldots,X_1)$, which measures the remaining uncertainty of the random process after observing the past trajectory. In the case of a stationary 1-$th$ order Markov process, the entropy rate has a simple form
\begin{eqnarray}
\mathcal{H}(\mathcal{X})&=&H(X_n|X_{n-1})\\\nonumber
&=&\sum_{i}(\pi_i \sum_{j}-P_{ij}\log P_{ij}).
\end{eqnarray}
Therefore, Site Entropy Rate (SER) is further defined to measure the influence of a node as follows:
\begin{equation}\label{equ:SER}
\text{SER}_i = \pi_i\sum_{j}-P_{ij}\text{log}P_{ij}
\end{equation}

SER can be divided into two parts: the stationary distribution term $\pi_i$ and the entropy term $\sum_{j}-P_{ij}\text{log}P_{ij}$. The $\pi_i$ tells the frequency at which a random walker visits webpage $i$; while, the entropy term $\sum_{j}-P_{ij}\text{log}P_{ij}$ measures the uncertainty of webpage $i$ jumping to the other webpages at one step. Consequently, SER measures the average information transmitted from webpage $i$ to the others in one step. The higher the SER value is, the more influential the webpage is. That is, the more likely the webpage is selected as a seed. Note that the concept of SER has been successfully used for image segmentation~\cite{liu-tuzel-TPAMI-entroy-rate-clustering-2014} and visual saliency detection~\cite{wang-wang-sailiency-cvpr10}.

Fig.~\ref{fig:SER} shows that SER measures the compactness and homogeneousness of a seed. Concretely, the central webpage in Fig.~\ref{fig:SER}(a) has the highest SER than the other ones in Fig.~\ref{fig:SER}. Moreover, compactness makes Fig.~\ref{fig:SER}(d) have a higher value than that of Fig.~\ref{fig:SER}(b).

Once the influence of each webpage is computed, a simple greedy method is proposed to find the influential seeds in Alg.~\ref{alg:generate-seeds}. Note that if the number of seeds are pre-determined, this problem can be converted as the maximal cover problem~\cite{Nemhauser-Wolsey-Fisher-max-cover-submodular-MP-78}, which could be efficiently solved by submodular optimization.

Alg.~\ref{alg:generate-seeds} has a complexity of $O(T\cdot nz(P)+N\log N+d\cdot N)$, where $T$ is number of iteration by power iteration~\cite{meyer-langville-pagerank-04} to solve $\pi$, and $nz(P)$ is number of non-zero elements in $P$. $O(T\cdot nz(P))$ is the complexity of computing $\pi$, $O(N\log N)$ is the time complexity of sorting webpages, and $O(d\cdot N)$ is the complexity of finding $d$ nearest webpages.

\begin{algorithm}[t!]
\SetAlgoLined
\caption{Selecting Influential Webpages as Seeds}\label{alg:generate-seeds}
\KwIn{A graph $G=(V,E,A)$, covering neighbors $d$}
\textbf{Initialize:} Seed webpages $\mathcal{S}\leftarrow \varnothing$\;
Compute SER of webpages by~\eqref{equ:SER}\;
Sort webpages by SER in a descent order, \emph{i.e.}, $\mathcal{SW} \leftarrow \text{sort}(\mathcal{X})$ \;
 \For{$\mathbf{x} \in \mathcal{SW}$}
 {
     $\mathbf{[x]}$ $\leftarrow$ Find $d$ nearest neighbor webpages of $\mathbf{x}$\;
     \If{the states of $\mathbf{x}$ and $\mathbf{[x]}$ are NOT \texttt{visited}}
     {
        $\mathcal{S}\leftarrow \mathcal{S}\cup \mathbf{x}$\;
        Label the states of $\mathbf{x}$ and $\mathbf{[x]}$ as \texttt{visited}\;
	 }
}
\KwOut{Seed webpages $\mathcal{S}$, and sorted webpages $\mathcal{SW}$.}
\end{algorithm}

\subsubsection{Simulating L\'{e}vy Walks by an Exploit-Explore Approach}

In this paper, we propose an Exploit-Explore (EE) approach to add a certain unexceptional flight, which simultaneously allocates a webpage to both the nearest the topic (Exploit) and the $topK$ nearest ones (Explore). Let $C_{\mathbf{s}}$ be a topic that has been grown from the seed $\mathbf{s}$. L\'{e}vy Walks-based Topic Generation (LWTG) is shown in Alg.~\ref{alg:simulate-LW-topic}.

\begin{algorithm}[t!]
\SetAlgoLined
\caption{Simulating L\'{e}vy Walks by Exploit-Explore}\label{alg:simulate-LW-topic}
\KwIn{A graph $G=(V,E,A)$, sorted webpages $\mathcal{SW}$, seed webpages $\mathcal{S}$, the number of multi-allocated topics $topK$}
\textbf{Initialize:} seed topic $C_{\mathbf{s}}\leftarrow \mathbf{s}$ ($\mathbf{s}\in \mathcal{S}$), threshold $th_{\mathbf{s}}=1.0$, and a set $\mathcal{C} \leftarrow \varnothing$\;

\For{each $\mathbf{x}$ $(\mathbf{x} \in \mathcal{SW})$}
{
    Compute $S_{\mathbf{x}\mathbf{s}}$ by~\eqref{eqt:webtopicsimilarity}\;
    $\{\widehat{C}_{\mathbf{s}}\} \leftarrow \text{Select}$ $topK$ nearest topics by sorting $S_{\mathbf{x}\mathbf{s}}$\;
    \For{each $\widehat{C}_{\mathbf{s}}$}
    {
        \eIf{$IsCut(\widehat{C}_{\mathbf{s}}, \mathbf{x}, th_{\mathbf{s}})$}
 		{
            $\widehat{C}_{\mathbf{s}} \leftarrow \widehat{C}_{\mathbf{s}} \cup \mathbf{x}$\;
		}
        {
             Put $\widehat{C}_{\mathbf{s}}$ into $\mathcal{C}$\;
            $th_{\mathbf{s}} \leftarrow \lfloor{\text{Avg}(\widehat{C}_\mathbf{s})\times 10}\rfloor\div 10$\;
		}
    }
}
\KwOut{a set of topics $\mathcal{C}$}
\end{algorithm}

Note that assigning a webpage to the $topK$ nearest topic not only makes the results stable and repeatable, but also adds the unexceptional flights. Therefore, the similarity between a candidate webpage $\mathbf{\widehat{x}}$ and a topic $C_{\mathbf{s}}$ is proposed as follows:
\begin{equation}\label{eqt:webtopicsimilarity}
S_{\mathbf{sx}} = \frac{\text{Avg}\big(\sum_{\mathbf{x} \in C_{\mathbf{s}}}(A(\mathbf{x,\widehat{x}})+A(\mathbf{\widehat{x},x}))\big)}{\text{Avg}(C_{\mathbf{s}})},
\end{equation}
where the Avg($\cdot$) function computes the mean affinity values in a set. The $S_{\mathbf{sx}}$ measures the relatively similarity between a topic and a webpage.

Following the similarity cascade~\cite{pang2013unsupervised}, the $IsCut(\cdot)$ function is proposed to generate the multi-granularity topics as follows:
\begin{equation}\label{eqt:is-accept}
IsCut(\cdot) = \left\{
\begin{array}{ll}
1, & \text{Avg}(C_{\mathbf{s}} \cup \mathbf{x}) < th_{\mathbf{s}},\\
0, & \text{otherwise},\\
\end{array}
\right.
\end{equation}
where $C_\mathbf{s} \cup \mathbf{x}$ means that a webpage $\mathbf{x}$ would be added into the topic $C_\mathbf{s}$. Because $\text{Avg}(C_\mathbf{s})$ is smaller than 1, the multi-thresholds $\lfloor{\text{Avg}(C_\mathbf{s})\times 10}\rfloor\div 10$ simply reduces the current threshold with a step size 0.1 in Alg.~\ref{alg:simulate-LW-topic}. For instance, when $\text{Avg}(C_\mathbf{s})$ is 0.76 and the current $th_{\mathbf{s}}$ is 0.8, $th_{\mathbf{s}}$ would be updated as 0.7.

\begin{figure}[t!]
  \centering
  \subfigure[A toy graph]{
    \label{fig:subfig:one-taxi} 
    \includegraphics[width=.15\textwidth]{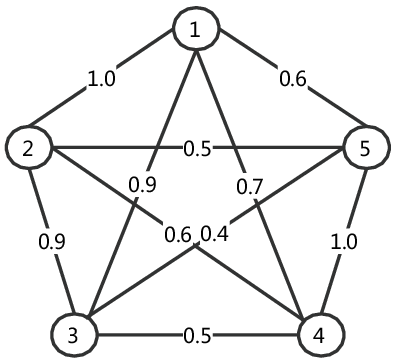}}
   \subfigure[HAC]{
    \label{fig:subfig:two-taxi} 
    \includegraphics[width=.27\textwidth]{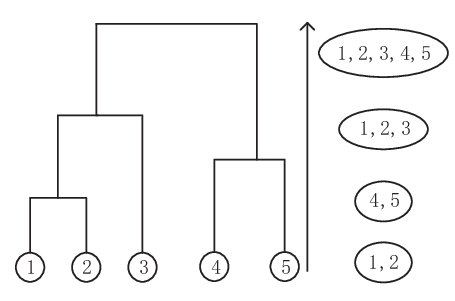}}
  \subfigure[LWTG]{
    \label{fig:subfig:two-taxi} 
    \includegraphics[width=.40\textwidth]{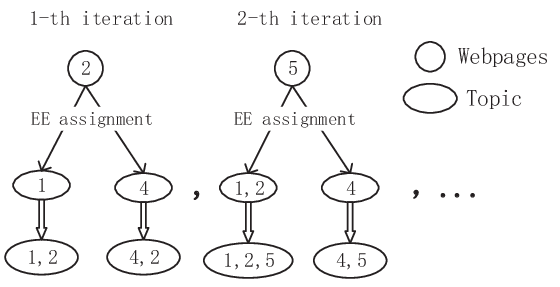}}
  \caption{ Comparison between HAC and LWT on a toy data. (b) Dendrogram generated by HAC and the corresponding clusters. (c) LWTG generates topics by the EE approach.}
  \label{fig:comparsion-HAC-LWT} 
\end{figure}

We would like to mention Hierarchical Agglomerative Clustering (HAC)~\cite{manning-book-08}, which is quite similar to LWTG at the first glance. Both of them discover topics in an agglomerative manner. However, they are different in terms of motive and technique: 1) HAC is originally designed for hierarchically clustering data; while LWTG aims at leveraging some exceptional flights in web topics rather than hierarchically clustering data; 2) the results of HAC depend on the selection of distance metrics, while LWTG adopts an EE method to generate web topics. Fig.~\ref{fig:comparsion-HAC-LWT} shows the results of HAC and LWTG, respectively.

The time complexity of \emph{generating LWTGs from seeds} is $O(N\cdot (|\mathcal{S}|+ topK \cdot num_{th}))$, where $|\mathcal{S}|$ is the number of seeds, and $num_{th}$ is the number of threshold $th_{\mathbf{s}}$. In practice, $|\mathcal{S}|$ is determined by the structure of graph, and $num_{th}$ is frequently smaller than 10.

\subsubsection{Generating L\'{e}vy Walks-based Topics across Multi-granularity Seeds}

\begin{algorithm}[t!]
\SetAlgoLined
\caption{Generating LWTG across Similarity Cascade}\label{alg:LW-topic-MS}
\KwIn{A graph $G=(V,E,A)$, a set of covering neighbors $D=\{2,\ldots,d,\ldots,4\}$, $k$-N$^2$ graph, the number of the multi-allocated topics $topK$}
\For{$d \in D$}
{
    Generate seeds from the $k$-N$^2$ via Alg.~\ref{alg:generate-seeds}\;
    Organizing topics from the graph $G$ via Alg.~\ref{alg:simulate-LW-topic}\;
}
\KwOut{a set of topics $\mathcal{C}$}
\end{algorithm}

In order to improve the recall of the topic detection system, Alg.~\ref{alg:LW-topic-MS} shows that generating LWTG across multi-granularity seeds consists of two steps, \emph{i.e.}, finding seed webpages by Alg.~\ref{alg:generate-seeds} and generating LWTGs from seed webpages by Alg.~\ref{alg:simulate-LW-topic}. Note that \emph{Generating LWTGs from seed webpages} dominates the time complexity of Alg.~\ref{alg:LW-topic-MS}. Therefore, the time complexity of Alg.~\ref{alg:LW-topic-MS} is $O( N\cdot(|\mathcal{S}|+topK \cdot num_{th}))$. In practice, the number of $\mathcal{S}$ is determined by a predefined value. Therefore, time complexity of Alg.~\ref{alg:LW-topic-MS} is nearly linear with respect to the number of webpages $N$. This makes Alg.~\ref{alg:LW-topic-MS} scale up to a large data set.

\begin{table*}[t!]
\footnotesize{
\begin{center}
\caption{A summary of datasets used in the experiments.}\label{tab:dataset}
\begin{tabular}{|m{0.1\textwidth}<{\centering}|c|m{0.06\textwidth}<{\centering}|m{0.1\textwidth}<{\centering}||m{0.06\textwidth}<{\centering}
|m{0.1\textwidth}<{\centering}||m{0.075\textwidth}<{\centering}|m{0.08\textwidth}<{\centering}|}
  \hline
  Data set& \#Topic & \#Webpage & \#Webpage in all topics & Dictionary size  & Average \#word$/$page &\# images in dataset & Average \# image$/$page \\  \hline \hline
MCG-WEBV & 73 & 3,660 & 832 & 9,212 & 35 & 108,925 & 29.8 \\\hline
YKS & 298 & 8,660 & 990 & 80,294& 228 & 71,063 & 8.2 \\\hline
\end{tabular}
\end{center}
}
\vspace{-4ex}
\end{table*}

\subsection{Topic Detection on Web by Poisson Deconvolution}

Once a similarity graph $G$ is built, a set of topics $C_k$ ($k=1,\ldots,K$) would be generated by Alg.~\ref{alg:LW-topic-MS}. PD estimates the weight of a topic $\mu_k$ under Poisson noise as follows~\cite{pang2013unsupervised}:
\begin{equation} \label{eqt:poissondeconvolution}
\begin{split}
w_{ij}&\sim \text{Poisson}(a_{ij})\\
s.t.: & \  \ w_{ij} = \sum_{k=1}^K \mu_k C_{k_{ij}},
\end{split}
\end{equation}
where $C_{k_{ij}}$ represents the edge between the $i$-th webpage and the $j$-th one. The interestingness of a topic is estimated as $i_k=\mu_k\cdot |C_k|$, where $|C_k|$ is the number of webpages in the topic $C_k$. In this work, PD is efficiently solved by a stochastic optimization~\cite{lin-pang-acclearte-PD-MMM19}.

\section{Experiment}\label{sec:experiments}

\textbf{Experiment Settings:} We evaluate our method on two public data sets, \emph{i.e.}, MCG-WEBV~\cite{Cao-Ngo-Zhang-Li-2011} and YKS~\cite{Zhang-Li-Chu-Wang-Zhang-Huang-2013}. MCG-WEBV was built from the ``Most viewed'' videos of ``This month'' on YouTube, the titles and the comments about video clips are represented as a set of words. YKS was downloaded from YouKu and Sina. To our best knowledge, there are no publicly accessible web topic datasets with nearly more than 96\% noisy webpages which do not belong to any topics. Both MCG-WEBV and YKS are very challenge to the approaches for clustering in a sea of noises.  

The texts of webpages are encoded with TF-IDF. The dot product is used to measure the textual similarity between webpages. The $k$ in the $k$-N$^2$ graph is 20 for MCG-WEBV. Since the contents of YKS are more complex and noisier than that of MCG-WEBV, the $k$ in the $k$-N$^2$G is assigned as 15 for the textual graph for YKS. The set of influence neighbors $D$ is $\{2,3,4\}$. The statistics of both data sets are summarized in Table~\ref{tab:dataset}

\textbf{Evaluation Criteria:} Top-10 $F_1$ versus Number of Detected Topics (NDT) and accuracy versus False Positive Per successfully detected Topic (FPPT)~\cite{pang2013unsupervised} are used as follows:

\begin{itemize}
\item \emph{Top-10 $F_1$ versus NDT:} A detected topic is matched with the ground truth, and then the top 10 $F_1$ scores are averaged to measure the performance:
\begin{eqnarray}\label{eqt:top-10-f1}
F_1=\frac{2\times Precision \times Recall}{Precision+Recall},
\end{eqnarray}
where $Precision$ is $\frac{|DT \cap GT|}{|DT|}$, $Recall$ is $\frac{|DT \cap GT|}{|GT|}$,
in which $DT$ is detected topic, $GT$ is ground truth topic, and $|\cdot |$ denotes the number of webpages in a topic.
\item \emph{Accuracy versus FPPT:} if a topic is correctly detected, FPPT is the number of false positive topics caused by a detection system. In this paper, accuracy is defined as follows:
\begin{equation}\label{equa:accuracy}
\text{Accuracy}=\frac{\# \text{Successful} }{\# \text{Groundtruth}},
\end{equation}
where $Successful$ means the detected topic $DT$ is successfully discovered, if Normalized Intersected Ratio (NIR) $\frac{|DT \cap GT|}{|DT \cup GT|}$ is larger than a threshold. Following the previous work~\cite{pang2013unsupervised}, 0.5 is used as the threshold of NIR in our experiments.
\item \emph{Scalability:} if the running time of an algorithm is changed from $T_M$ to $T_N$ when the size of the testing data set is increased from $M$ to $N$ ($N\geq M$), the scalability of the algorithm is defined as follows:
    \begin{equation}
    \text{Scalability} = \frac{M\cdot T_N}{N\cdot T_M}.
    \end{equation}
 Obviously, if the time complexity of an algorithm is linear with respect to the number of samples, scalability should be 1; otherwise, the time complexity of an algorithm is either sublinear or superlinear, when scalability is less than 1 or is larger than 1. Therefore, the smaller scalability is, the more scalable an algorithm is.
\end{itemize}

\textbf{Baselines and Alternative Approaches:}
Our experimental goals include two aspects:

\uppercase\expandafter{\romannumeral1}) Compare the proposed LWTG with two state-of-the-art methods to cluster in a sea of noises:
\begin{itemize}
\item [1.] \textbf{Robust Spectral Clustering (RSC) for noisy data~\cite{bojchevski-RSC-kdd-2017}.} This paper handled noises by decomposing a spare and latent graph. However, this method assumes that noise data are sparse. In contrast, in our topic detection on web scenario, about 95\% data are the noise webpages.
\item [2.] \textbf{Skinny-Dip (SD)~\cite{maurus-plant-skinny-dip-kdd-16}.} SD handled abundant noises based on Hartigan's elegant dip test of unimodality. That is, SD recursively finds clusters based on dips into univariate projections of data. Therefore, very high-dimensional features from webpages make SD inefficient.
\end{itemize}
Note that RSC and SD are not intended to detect topics on web, but to cluster from noisy data. The comparisons between RSC, SD and LWTG are intended to verify whether these clustering algorithms could efficiently detect topics on web. In the following experiments, for RSC, the number of clusters is assigned as the number of ground truth topics, \emph{i.e.}, 73 for MCG-WEBV and 298 for YKS, as shown in Table.~\ref{tab:dataset}; while for SD, the number of clusters is automatically determined by the algorithm in~\cite{maurus-plant-skinny-dip-kdd-16}.

\uppercase\expandafter{\romannumeral2}) Compare the proposed approach with three state-of-
the-art methods for topic detection on web:
\begin{itemize}
\item [1.]\textbf{Multi-Modality Graph (MMG)~\cite{Zhang-Li-Chu-Wang-Zhang-Huang-2013}.} Zhang \emph{et al.}~\cite{Zhang-Li-Chu-Wang-Zhang-Huang-2013} used both the Near Duplicated Keyframes (NDKs) of video clips and the text information to build the similarity graph~\cite{papadopoulos-11}. Moreover, graph shift~\cite{liu-icml-2009} was used to discover hot topics from the similarity graph.
\item [2.]\textbf{PD with  Non-negative Matrix Factorization on Graph (NMFG)~\cite{pang-tao-neurocomputing-2018}.} The only difference between NMFG and LWTG is the way to generate topics for PD. NMFG~\cite{pang-tao-neurocomputing-2018} utilized the Non-negative Matrix factorization on Graph (NMG)~\cite{yang2012clustering} to generate multi-granularity topics. Note that NMG is a time-consuming approach to generate clusters for noiseless data.
\item [3.]\textbf{Latent Poisson Deconvolution (LPD)~\cite{pang-tao-lpd-icme-2016}.} This method achieves the state-of-the-art performances on both MCG-WEBV and YKS. Rather than using a single $k$-N$^2$ graph in PD, LPD leverages multiple $k$-N$^2$ graphes to rank topics. This demonstrates that our approach can achieve acceptable results without exploiting multiple \mbox{$k$-N$^2$} graphes.
\end{itemize}

\subsection{Evaluation of LWTG}

\subsubsection{Effectiveness of the EE selection}

\begin{figure}[t!]
\centering
\subfigure[]{
\label{fig:sub:MCG-EE-accuracy}
\includegraphics[width=.4\textwidth]{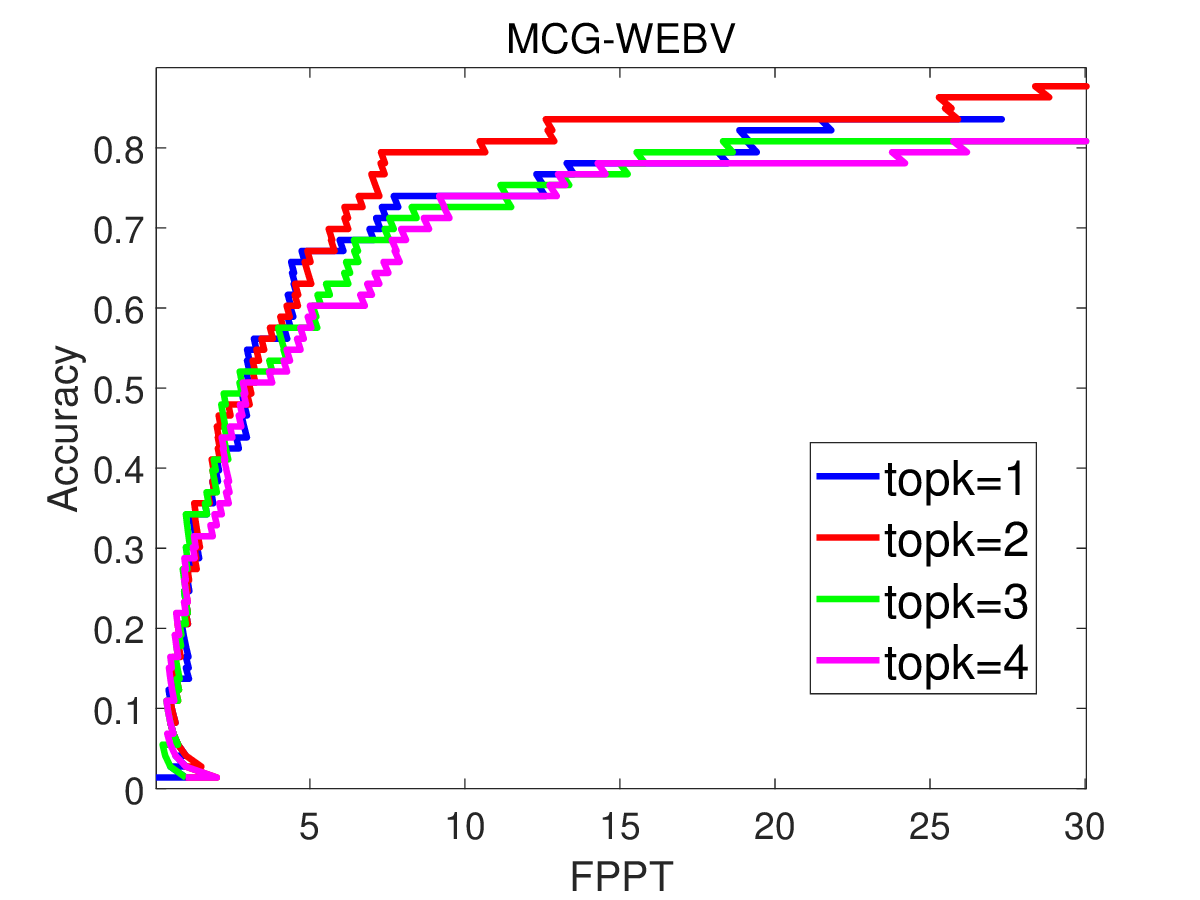}}
\subfigure[]{
\label{fig:sub:MCG-EE-f1}
\includegraphics[width=.4\textwidth]{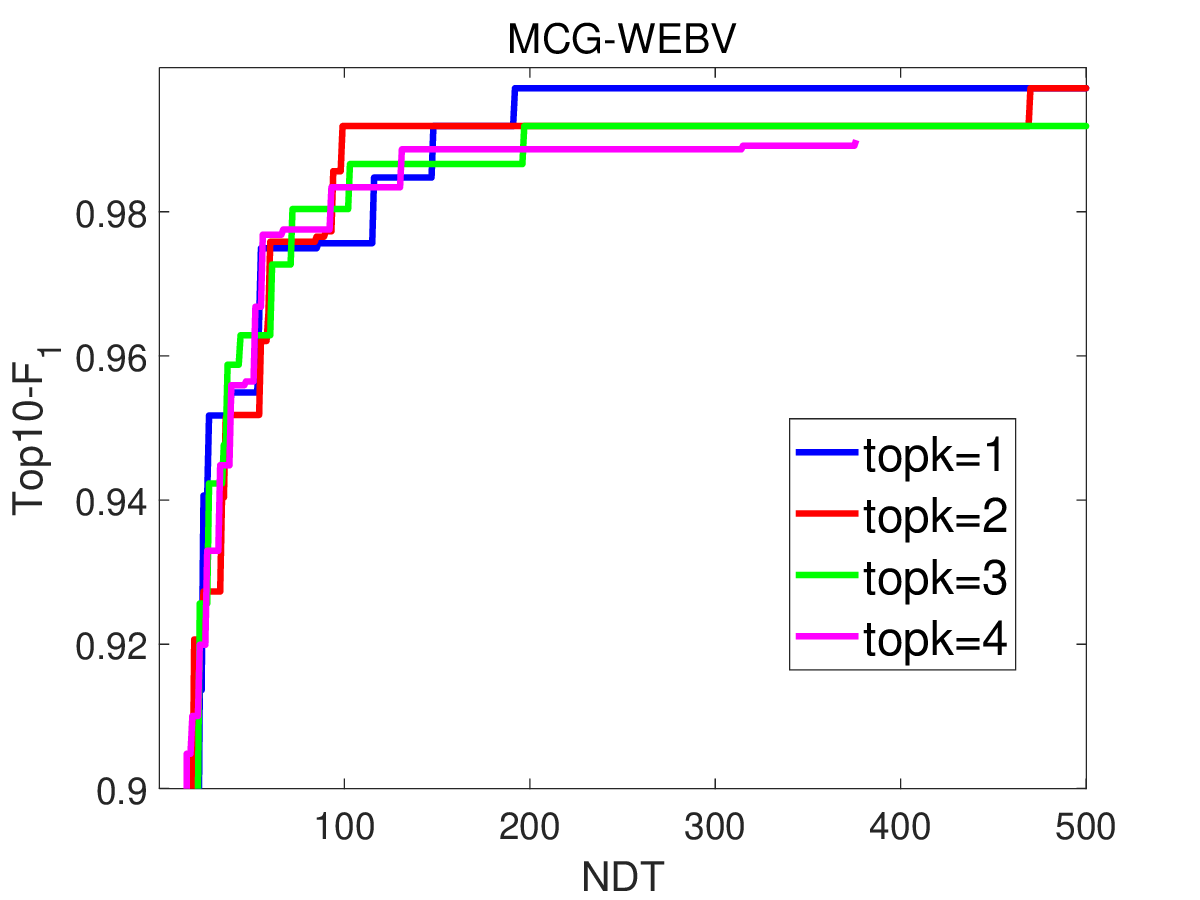}}
\caption{Effectiveness of the EE selection in simulating l\'{e}vy walks on MCG-WEBV (best viewed in color).}\label{fig:effectiveness-EE-MCG}
\end{figure}

In this subsection, we verify the effectiveness of the proposed EE selection. As shown in Fig.~\ref{fig:sub:MCG-EE-accuracy}, a certain of randomness (\emph{i.e.}, $k{=}2$) obtains a higher accuracy than a fixed approach (\emph{i.e.}, $k{=}1$). For instance, when FPPT is equal to 10, $topK=2$ significantly outperforms $topK{=}1$, \emph{i.e.}, 0.8 v.s. 0.72. As expected, when more randomness (such as $topK{=}3$, $topK{=}4$) is added into the generation of topics, the number of accurate detected hot topics would be decreased. These empirically results indicate the following observation:~\emph{Too many or too small number of unexceptional flights would damage the structure of a web topic.} The explantation is that the setting $topK{=}1$ always selects the short ``flights''; in contrast, the setting $topK{=}4$ injects too many the long ``flights''.

Interestingly, Fig.~\ref{fig:sub:MCG-EE-f1} further shows that the top 10 $F_1$ scores of the different settings are nearly similar to each other. Because only 10 best matched topics are selected from about 2,500 candidates during the evaluation. It indicates that~\emph{different topics should be injected with the different number of the unexceptional flights.} Interestingly, if we compare Fig.~\ref{fig:sub:MCG-EE-accuracy} with Fig.~\ref{fig:sub:MCG-EE-f1}, the optimal number of unexceptional flights is unanimously same, \emph{i.e.}, $topK{=}2$.

In the following experiments, $topK$ is assigned as $2$. Because $topK{=}2$ makes LWTG balance well between two evaluation criteria.

\subsubsection{Quantitative Comparisons with the State-of-The-Art
Methods}

\begin{figure}[t!]
\centering
\subfigure[]{
\label{fig:sub:MCG-accuracy-clusterings}
\includegraphics[width=.23\textwidth]{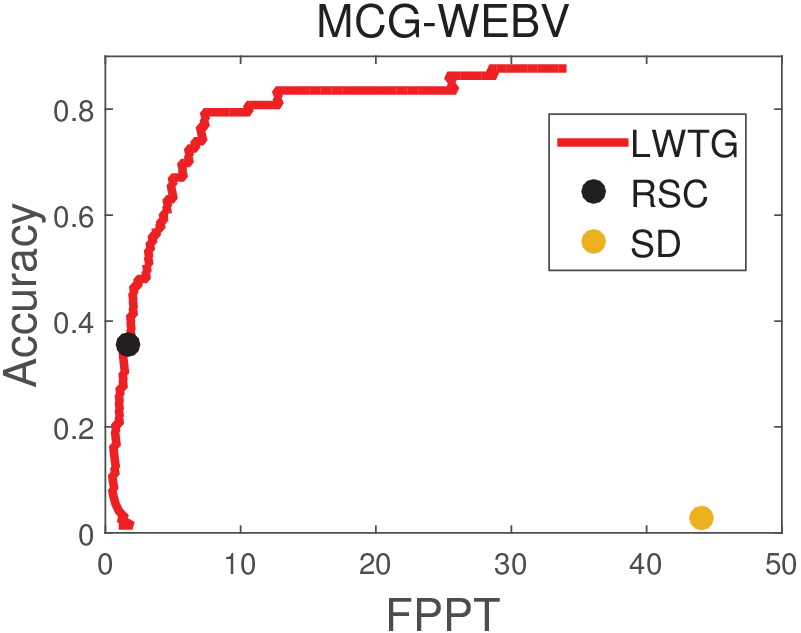}}
\subfigure[]{
\label{fig:sub:MCG-top-f1-clusterings}
\includegraphics[width=.23\textwidth]{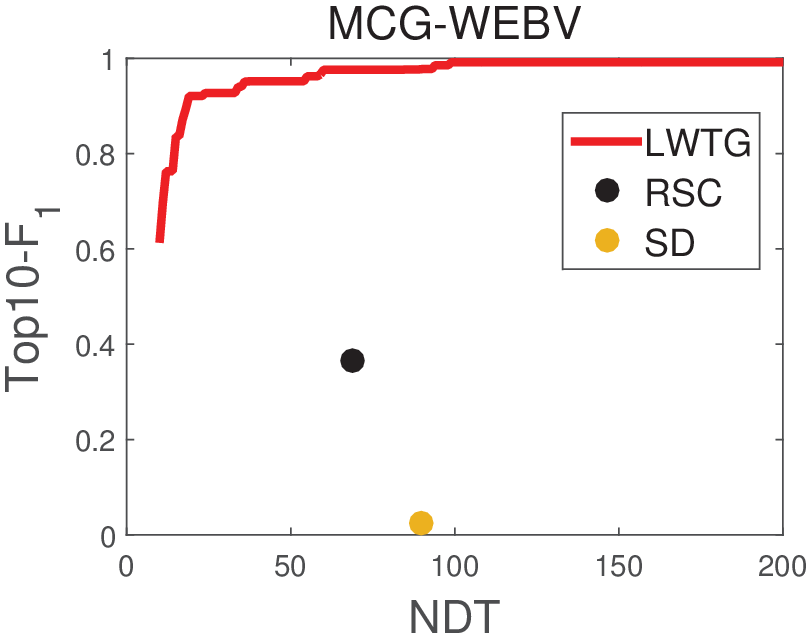}}
\caption{The comparisons between RSC, SD and our method on MCG-WEBV (best viewed in color).}\label{fig:comparison-clustering-MCG}
\end{figure}

Fig.~\ref{fig:comparison-clustering-MCG} illustrates that our method significantly outperforms SD and RSC. As expected, SD achieves nearly zero Accuracy in Fig.~\ref{fig:sub:MCG-accuracy-clusterings} and zero top-10 $F_1$ in Fig.~\ref{fig:sub:MCG-top-f1-clusterings}. Because the dimension of TF-IDF is extremely high for SD, for instance, the dimension of TF-IDF extracted from MCG-WEBV is 9,212. As a result, SD inefficiently discovers clusters from the high-dimensional and noisy features. Interestingly, the Accuracy of RSC in Fig.~\ref{fig:sub:MCG-accuracy-clusterings} is comparable to our method; while, top-10 $F_1$ of RSC in Fig.~\ref{fig:sub:MCG-top-f1-clusterings} is much worse than our method. The explanation is that the quality of clusters generated by RSC is significantly corrupted by the abundant noises. As a comparison, our method achieves competitive performances over two evaluation criteria without the complex model and its optimization.

\begin{figure}[h!]
\centering
\subfigure[]{
\label{fig:sub:YKS-accuracy-clusterings}
\includegraphics[width=.23\textwidth,height=0.18\textwidth]{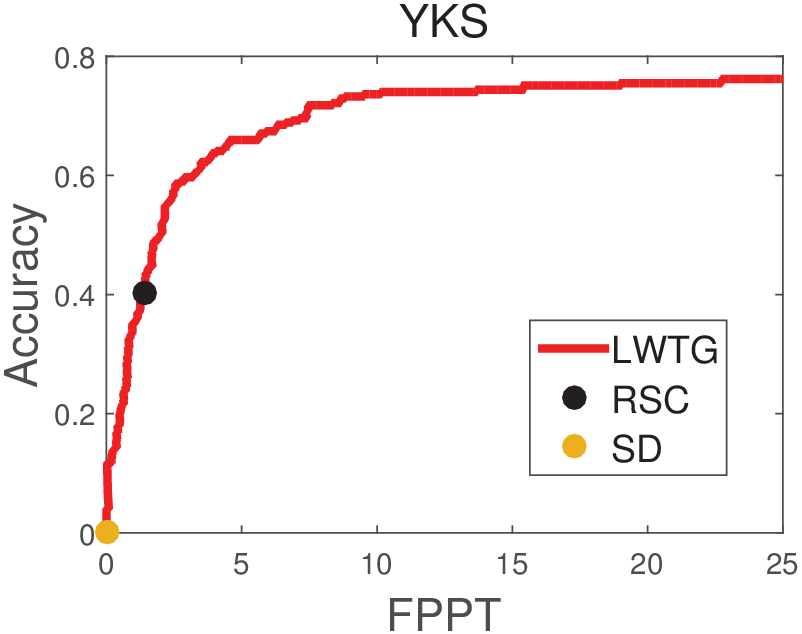}}
\subfigure[]{
\label{fig:sub:YKS-top-f1-clusterings}
\includegraphics[width=.23\textwidth,height=0.18\textwidth]{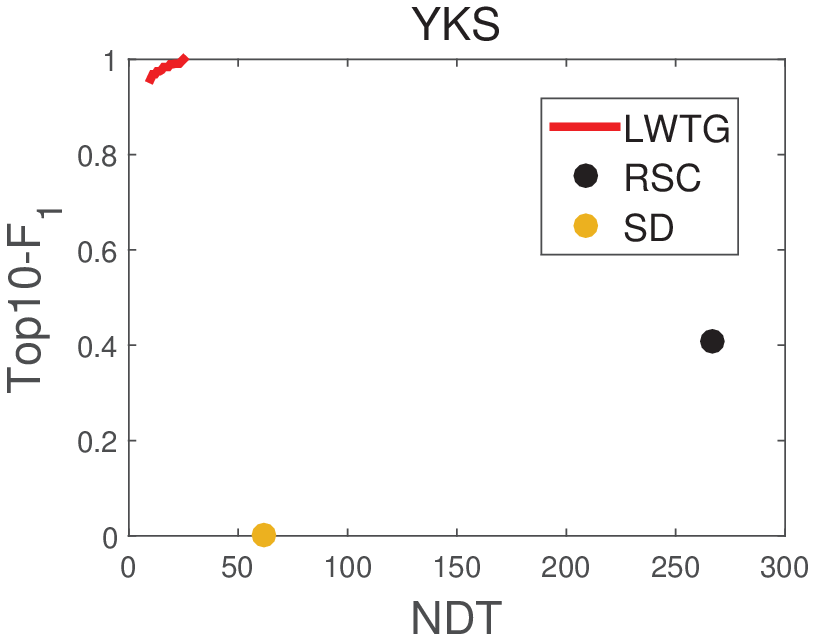}}
\caption{The comparisons between RSC, SD and our method on YKS (best viewed in color).}\label{fig:comparison-clustering-YKS}
\end{figure}

Fig.~\ref{fig:comparison-clustering-YKS} further illustrates the comparisons between RSC, SD and our method on YKS. Note that the performance of SD is not plotted in Fig.~\ref{fig:comparison-clustering-YKS} since the very high-dimension of YKS (\emph{i.e.}, 80,294) makes SD incapable to generate any clusters. Fig.~\ref{fig:comparison-clustering-YKS} shows that our method consistently outperforms RSC and SD on YKS.

In summary, in terms of effectiveness, the proposed LWTG achieves both the highest Accuracy value and the highest top-10 $F_1$ score than those of RSC and SD, when the same number of topics are generated for different methods.

Table.~\ref{tbl:runing-time-of-denoise-clustering} further compares the running time of RSC~\cite{li-liu-chen-tang-NRSC-iccv-07}, SD~\cite{maurus-skinny-dip-kdd16} and our method. Although, in terms of Accuracy \emph{v.s.} FPPT, our method achieves a very similar accuracy to that of RSC in Figs.~\ref{fig:sub:MCG-accuracy-clusterings} and~\ref{fig:sub:YKS-accuracy-clusterings}, the recall of our method significantly surpasses RSC. For instance, Table.~\ref{tbl:runing-time-of-denoise-clustering} shows that the recall of LWTG is 0.88 on MCG-WEBV; in contrast, only 35\% hot topics are found by RSC. The explanation is that the approach of simulating L\`{e}vy walks nature of topics guarantees both the recall and precision of a detection system.

Moreover, one interesting observation is that when the number of webpages is increased from 3,660 to 8,660, the scalability of RSC is $6.44$, costing more than 15$\times$ (\emph{i.e.},$125/8$) time on YKS as long as that on MCG-WEBV. In contrast, the scalability of our method is $1.32$, only increasing about 3$\times$ (\emph{i.e.},$233/73$) time on YKS as long as that on MCG-WEBV. The explanation is that RSC has the time complexity $O(\gamma\cdot |E|)$, where $\gamma$ is the number of unique edge weights per node~\cite{bojchevski-RSC-kdd-2017}. In our scenario, $\gamma$ is almost equal to $k$, and $|E|$ is usually a quadratical webpages $N$. In contrast, the time complexity of our method, $O( N\cdot(|\mathcal{S}|+topK \cdot num_{th}))$, is nearly linear with respect to $N$.

\begin{table}[t!]
\small{
    \caption{Comparisons of running time (in seconds) and accuracy of a system (in~\eqref{equa:accuracy}) on a PC with 3.6 Hz CPU with 32 G memory.}
    \label{tbl:runing-time-of-denoise-clustering}
    \centering
    \begin{tabular}{|m{0.10\textwidth}<{\centering}|c|m{0.1\textwidth}<{\centering}|m{0.1\textwidth}<{\centering}|}
        \hline
        Dataset (\#webpages) & LWTG (Recall) & RSC (Recall) & SD (Recall)\\
        \hline
        \hline
        MCG-WEBV (3,660) & $73 (\mathbf{0.88})$ & $\mathbf{8} (0.35)$ & $20 (0.02)$\\
        \hline
        YKS \qquad \qquad (8,660) & $233 (\mathbf{0.76})$ & $\mathbf{125} (0.40)$ & $152 (0.00)$\cr \hline\hline
        Scalability  & $\textbf{1.32} $ & $6.44$ & $3.15 $\cr \hline
    \end{tabular}
    }
\end{table}

\subsection{Evaluation of Topic Detection on Web}

In this subsection, we compare the proposed approach on the benchmark data sets. To make comparisons as fair as possible, we use the same experimental settings proposed by each data set.

\subsubsection{\textbf{Web-Video Topic Detection in MCG-WEBV}}

\begin{table*}[t!]
  \centering
  \caption{Comparisons of running time (in seconds) among different methods on a PC with 3.6 Hz CPU with 32 G memory.}
  \label{tbl:runing-time-of-topic-detection}
    \begin{tabular}{|m{0.10\textwidth}<{\centering}|m{0.1\textwidth}<{\centering}|m{0.1\textwidth}<{\centering}|m{0.1\textwidth}<{\centering}|m{0.1\textwidth}<{\centering}|}
    \hline
    Dataset (\#webpage) & LWTG (\#topic) & LPD (\#topic) & NMFG (\#topic) & MMG (\#topic)\\
    \hline
    MCG-WEBV (3,660)& 73 (2,504) &11,545 (7,685) &5,248 (4,238) & \textbf{15} (430) \cr\hline
    YKS \qquad \qquad (8,660)& \textbf{233} (7,458) & 74,812 (7,524) & 28,024 (5,714) & 252 (445) \cr\hline\hline
    Scalability & $\textbf{1.32}$ & $2.68 $ & $2.21$ &
    $6.97$   \cr\hline
\end{tabular}
\end{table*}

\begin{figure}[t!]
\centering
\subfigure[Accuracy versus FPPT]{
\label{fig:sub:MCG-state-of-the-art-accuracy-fppt}
\includegraphics[width=.35\textwidth]{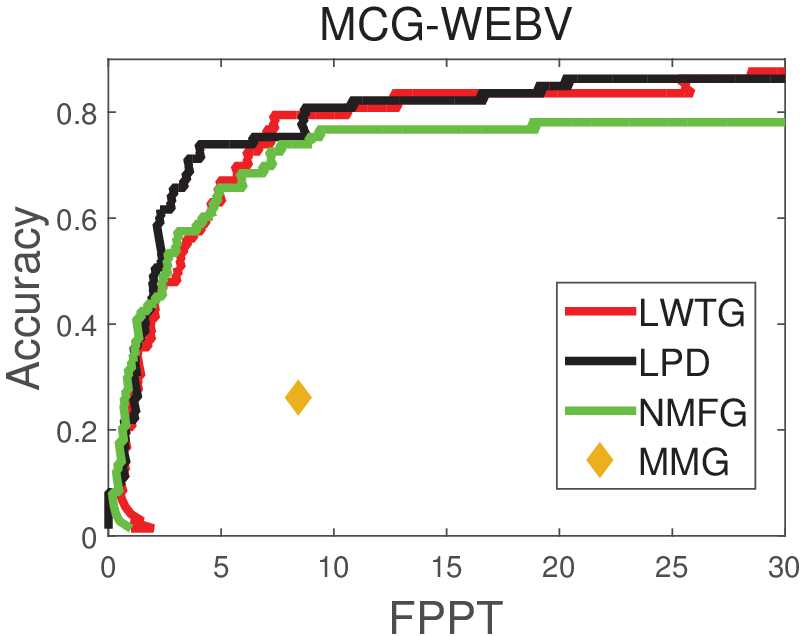}}
\hspace{.8pt}
\centering
\subfigure[Top-10 $F_1$ versus NDT]{
\label{fig:sub:MCG-state-of-the-art-f1-NDT}
\includegraphics[width=.36\textwidth]{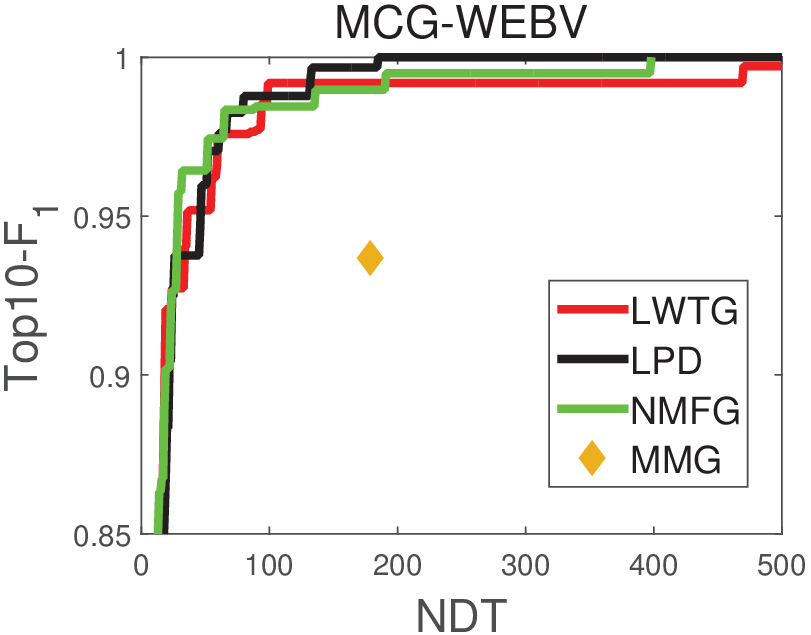}}
\caption{Comparisons between the state-of-the-art methods and our method on MCG-WEBV (best viewed in color).}\label{fig:state-of-the-art-MCG}
\end{figure}

Fig.~\ref{fig:sub:MCG-state-of-the-art-accuracy-fppt} shows the comparison results in terms of Accuracy \emph{v.s.} FPPT on MCG-WEBV. There are two observations in Fig.~\ref{fig:state-of-the-art-MCG} as follows:
\begin{itemize}
\item The highest accuracy of LWTG is comparable to that of LPD~\cite{pang-tao-lpd-icme-2016}, but significantly outperforms that of \mbox{NMFG}~\cite{pang-tao-neurocomputing-2018}, when FPPT is larger than 8 in Fig.~\ref{fig:sub:MCG-state-of-the-art-accuracy-fppt}. Note that LPD uses the time-consuming but high-efficient NMG~\cite{yang2012clustering} to generate topics from two $k$-N$^2$ graphes. In contrast, the proposed \emph{model-free} and \emph{optimization-free} approach avoids to design a complex model and its optimization.
\item The number of hot topics successfully discovered by LWTG is slightly less than that of LPD. The main explanation is that LPD uses two $k$-N$^2$ graphes to generate almost 7,600 topics on MCG-WEBV; while, the proposed LWTG only generates 2,500 topics.
\end{itemize}

Fig.~\ref{fig:sub:MCG-state-of-the-art-f1-NDT} illustrates that LWTG is comparable to the other state-of-the-art methods~\cite{pang-tao-neurocomputing-2018}~\cite{pang-tao-lpd-icme-2016}, when
the number of detected topics is about 100. It means that LWTG at least accurately generates best matched top-10 hot topics. However, when NDT ranges from 100 to 400, LWTG is slightly worse than NMFG and LPD. The explantation is that some inaccurate topics are wrongly ranked ahead of the accurate ones, as observed in Fig.~\ref{fig:sub:MCG-state-of-the-art-accuracy-fppt}.

In terms of efficiency, Table~\ref{tbl:runing-time-of-topic-detection} shows that the running time of NMFG~\cite{pang-tao-neurocomputing-2018} and LPD~\cite{pang-tao-lpd-icme-2016} are respectively about 72$\times$ (\emph{i.e.}, 5,248/73) and 158$\times$ (\emph{i.e.}, 11,545/73) as long as that of LWTG. It means that our method is very promising to scale up to a large-scale data set; meanwhile, the effectiveness of LWTG is comparable to the state-of-the-art methods.

\begin{figure}[h!]
\centering
\subfigure[Accuracy versus FPPT]{
\label{fig:sub:YKS-state-arts-accuracy}
\includegraphics[width=.4\textwidth]{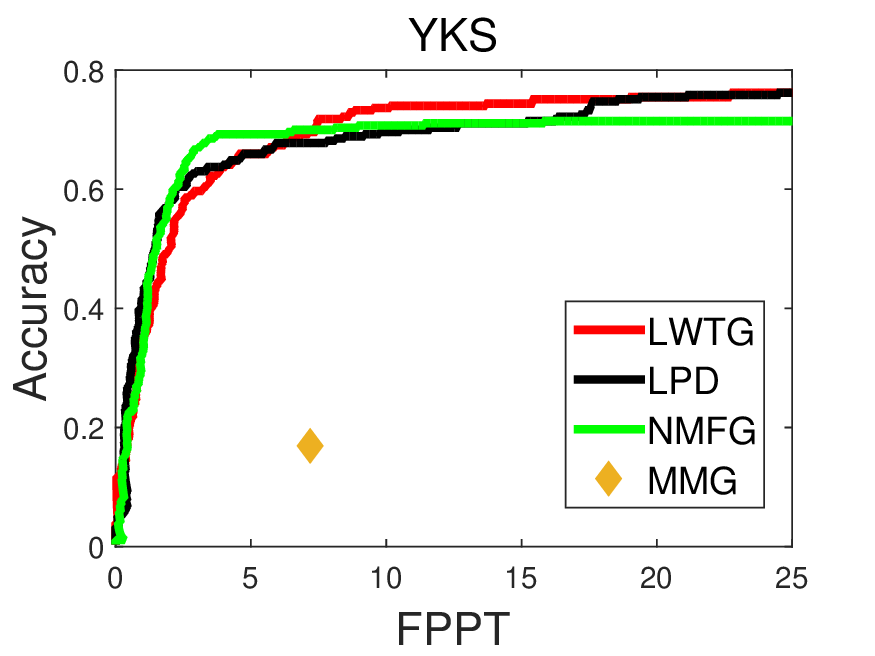}}
\subfigure[Top-10 $F_1$ versus NDT]{
\label{fig:sub:YKS-state-arts-f1}
\includegraphics[width=.4\textwidth]{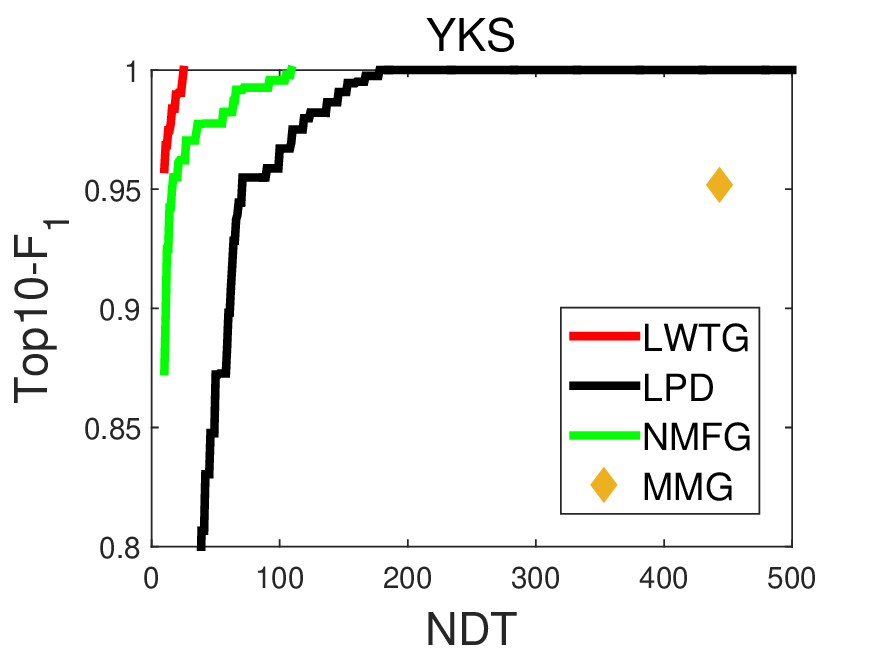}}
\caption{Comparisons between the state-of-the-art methods and our method on YKS (best viewed in color).}\label{fig:YKS-state-arts}
\end{figure}

\subsubsection{\textbf{Web Topic Detection in YKS}}

YKS, a cross-platform data set, requires to grasp more diverse types of topics than MCG-WEBV. Therefore, YKS is more challenging than MCG-WEBV.

Fig.~\ref{fig:sub:YKS-state-arts-accuracy} illustrates the accuracy \emph{v.s.} FPPT curves on YKS. Accuracy of our approach consistently outperforms these state-of-the-art methods when FPPT is larger than 8 in Fig.~\ref{fig:sub:YKS-state-arts-accuracy}. For instance, our method achieves 0.76 Accuracy, outperforming NMFG~\cite{pang-tao-neurocomputing-2018} about 0.05. Although the accuracies of our method are worse than that of NMFG~\cite{pang-tao-neurocomputing-2018} when FPPT ranges from 0 to 8, our method quickly surpass NMFG when FPPT is larger than 8.

Fig.~\ref{fig:sub:YKS-state-arts-f1} shows that our method consistently outperforms NMFG~\cite{pang2013unsupervised}, LPD~\cite{pang-tao-lpd-icme-2016} and MMG~\cite{Zhang-Li-Chu-Wang-Zhang-Huang-2013}, if the same number of topics is generated. For instance, Top-10 $F_1$ of our method is 1, while MMG~\cite{Zhang-Li-Chu-Wang-Zhang-Huang-2013} is 0.95, when 445 topics are generated for both methods. This demonstrates the generalization ability of our approach across different data sets.

In terms of efficiency, Table~\ref{tbl:runing-time-of-topic-detection} shows that the running time of NMFG~\cite{pang-tao-neurocomputing-2018} and LPD~\cite{pang-tao-lpd-icme-2016} are respectively about 120$\times$ (\emph{i.e.}, 28,024/233) and 321$\times$ (\emph{i.e.}, 74,812/233) as long as that of LWTG. It means that our method consistently scales up for a large-scale data set. It should note that the running time of MMG~\cite{Zhang-Li-Chu-Wang-Zhang-Huang-2013} is greatly increased by the number of webpages. For instance, the scalability of MMG is 6.97; while, the scalability of the proposed LWTG is 1.32.

\section{CONCLUSION}\label{sec:conclusion}

In this paper, we have described a method to efficiently generate topics by simulating L\`{e}vy walk nature of web topics, leading to the results being comparable to the state-of-the-art methods. More importantly, the proposed LWTG is more scalable than the state-of-the-art methods, shining a new light on the scalability of topic detection on web. There are significant distinctions between the proposed method and the previous studies in generating web topics:
\begin{itemize}
\item We demonstrate the statistically similar features between L\`{e}vy walks and web topics in the similarity space;
\item The proposed LWTG balances well between the effectiveness of the PD-based method~\cite{pang2013unsupervised} in achieving the high performances and the efficiency of the model-free approach towards scaling up to a large-scale data set;
\item The proposed LWTG assumes no prior information about web topics, except the assumption of simulating exceptional long ``flights'', in contrast to the model-based methods~\cite{pang-tao-lpd-icme-2016}~\cite{pang-tao-neurocomputing-2018} which often involve a time-consuming optimization.
\end{itemize}

The promising results of this paper motivate a further examination of the proposed LWTG. Firstly, more effective methods about measuring the similarity between a webpage and a topic may bring more accurate topics over~\eqref{eqt:webtopicsimilarity} used here. Secondly, evaluating the interestingness of topics in a dynamic approach may further accelerate LWTG for the large-scale data set.

\section*{References}

\bibliography{icme,tmmtopic}

\end{document}